# Imaging with the invisible light

R. Bellazzini[a], G. Spandre[a*], A. Brez[a], M. Minuti[a], L. Baldini[a], L. Latronico[a], M.M. Massai[b], N. Omodei[a], M. Pesce-Rollins[a], C Sgrò[a], M. Razzano[a], M. Pinchera[a], J. Bregeon[a], M. Kuss[a], A. Braem[c]

[a]*Istituto Nazionale di Fisica Nucleare di Pisa, Largo B. Pontecorvo, 3 I-56127 Pisa, Italy*
[b]*University of Pisa and INFN-Pisa, Largo B. Pontecorvo, 3 I-56127 Pisa, Italy*
[c]*CERN, CH-1211 Gene`ve 23, Switzerland*

**Abstract**

We describe a UV photo-detector with single photon(electron) counting and imaging capability. It is based on a CsI photocathode, a GEM charge multiplier and a self triggering CMOS analog pixel chip with 105k pixels at 50 μm pitch. The single photoelectron produced by the absorption of a UV photon is drifted to and multiplied inside a single GEM hole. The coordinates of the GEM avalanche are reconstructed with high accuracy (4 μm rms) by the pixel chip. As a result the map of the GEM holes, arranged on a triangular pattern at 50μm pitch, is finely imaged.

**1. Introduction**

The idea of position sensitive gas detectors with CsI photocathodes was introduced and developed by Charpak's group [1] at the end of the 80s when the first parallel-plate chamber with a CsI photocathode was successfully tested. Soon afterward solid photocathodes were coupled to wire chambers [2] and, with the introduction of new detector construction technologies as photo-lithography and advanced PCB, to new types of charge amplification structures such as MSGC[3,4,5], MGC[6], GEM[7,8,9].
Since the beginning, these types of detectors found great interest and widespread applications in several, different fields, from high energy physics, to nuclear physics, astrophysics, industry, plasma diagnostic and medical imaging and for this reason, in the last two decades, a large variety of "*gaseous photomultipliers*" have been developed and successfully tested. In the development of photon detectors the main requests are for detection efficiency, particularly to single photon, localization accuracy and time stability. The use of GEMs in gaseous photodetectors with CsI photocathode, has allowed, due to the shadowing effect provided by the GEM itself, to efficiently reduce the avalanche-induced photon and ion feedback which spoils the performance and limits the gain and the lifetime of the detector.
In this work we describe a UV photodetector based on a single GEM charge multiplier coupled to a custom CMOS VLSI pixel array which is, at the same time, the charge collecting electrode and the analog, self triggering, read-out electronics. The fine pitch of both the read-out system and of the amplification structure (50 μm) and the possibility to work in single photon detection allows to reach avalanche reconstruction accuracy of 4 μm.

**2. The UV photon detector**

Two different photocathode configurations have been studied: the so-called *semitransparent* mode (fig.1, left panel), obtained by evaporating ~100 Å CsI film onto the quartz window that closes the gas volume of the detector and the *reflective* mode (fig.1, right panel) in which the UV photon converter layer (thousands Å) is deposited onto the top side of the GEM. With this last method the photoelectrons extracted from the CsI are focused, by the electric field at the top GEM surface, into the nearest GEM hole and here multiplied. The

*reflective* approach allows the use of thick photocathodes which are easier to handle and have higher quantum efficiency (10-20% at 150 nm) compared to the semitransparent ones but, on the other hand, a photo-detector operating in this configuration needs a special gold pre-coating on the GEM which introduces a lower geometrical efficiency (~50% in our case) and a lower gas gain.

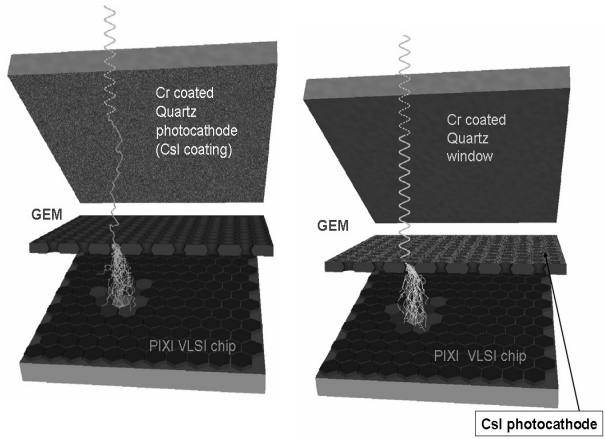

Fig. 1 - Schematic view of the detector in the semitransparent (left panel) and reflective (right panel) configuration.

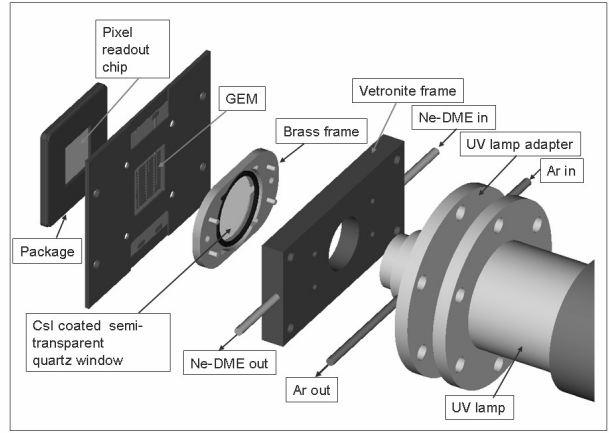

Fig. 2 – The UV photo-detector with single electron sensitivity.

With respect to the reflective mode of operation the detector with semitransparent photocathode is simpler to fabricate, has a higher gas gain and geometrical efficiency but the low thickness of the CsI coating prevents to obtain high Q.E. Fig. 2 shows a sketch view of the detector assembly and the UV-lamp adapter.

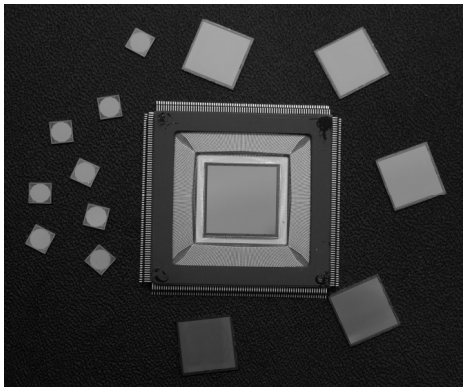

Fig. 3 – The three CMOS VLSI chips in comparison. The last version with 105.600 pixels is shown bonded to its ceramic package (304 pins).

The brass frame in fig.2 houses the quartz window and defines the 1 mm drift region. The transfer gap between the GEM bottom and the charge collection pads of the VLSI ASIC has also 1 mm thickness. This chip is the third generation [10,11] of a series of custom CMOS ASICs of increasing complexity, larger size, reduced pitch and improved functionality (fig.3). In the top metal layer of the chip, 105600 hexagonal pixels are arranged at 50 µm pitch in an 300×352 squared matrix of $15 \times 15 mm^2$ active area (corresponding to a pixel density of $470/mm^2$). A full electronics chain (pre-amplifier, shaping amplifier, sample and hold, multiplexer) is integrated under each pixel in the remaining metal, polysilicon and oxide layers of the 0.18 µm CMOS technology. The chip integrates more than 16.5 million transistors and it is subdivided in 16 identical clusters of 6600 pixels (22 rows of 300 pixels) or alternatively in 8 clusters of 13200 pixels (44 rows of 300 pixels) each one with an independent differential analog output buffer.

A new feature implemented in the chip is a customizable internal self-triggering capability with independently adjustable thresholds for each cluster. Groups of 4 contiguous pixels (mini-cluster) contribute to a local trigger with a dedicated amplifier whose shaping time ($T_{shaping} \sim 1.5$ µs) is roughly a factor of two faster than the shaping time of the analog charge signal. In self-trigger operation the read-out time and the amount of data to be transferred result vastly reduced (at least by a factor 100) with respect to the standard sequential read-out mode of the full matrix (which is still available, anyway). This is due to the relatively small number of pixels (400-500) within the region of interest. The main characteristics of the VLSI chip are summarized in Table 1.

To exploit the high granularity of the read-out plane, a non-standard gas amplification structure (GEM) with a pitch of 50 µm that matches the read-out pitch has been manufactured at CERN. Optimal spatial resolution and imaging capability have been obtained. The technological challenge in the fabrication of this type of GEM has been the precise and uniform etching of the very narrow biconical multiplication holes, only 33 µm and 15 µm diameter respectively at the top and in the middle of the kapton layer. With this gas multiplier, large effective gains at voltages at least 70V less than in standard 90 µm GEM have been reached. A gas gain of around 1000 has been obtained in 50% Ne – 50% DME at 450V across the GEM. This is likely due to the higher field lines density inside the very narrow amplification holes. A photo of the detector ready to be mounted on the control mother-board is shown in fig. 4.

Table 1
**Main ASIC characteristics:**

| | |
|---|---|
| Peaking time (externally adjustable) | 3÷10 µs |
| Full scale linear range | 30000 electrons |
| Pixel noise | 50 electrons ENC |
| Read-out mode | asynchronous and synchronous |
| Trigger mode | Internal, external, self-trigger |
| Read-out clock | Up to 10 MHz |
| Self-trigger threshold | 2200 electrons (10% FS) |
| Frame rate | Up to 1 KHz (event window self trigger) |
| Parallel analog output buffer | 1, 8 or 16 |
| Access to pixel counter | direct (single pixel) or serial (8-16 clusters, full matrix, region of interest) |
| Fill fraction (metal to active area ratio) | 92% |
| Total power dissipation | ~ 0.5 Watt |

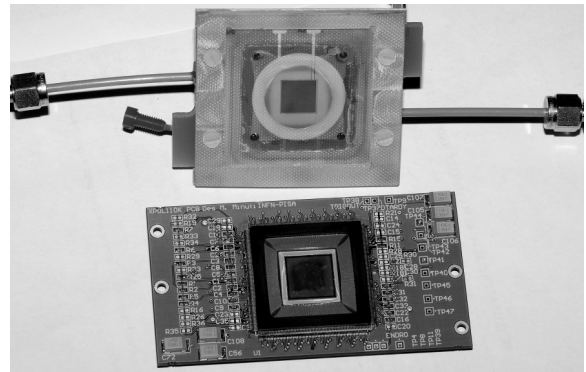

Fig. 4 - Detector and VLSI chip ready to be assembled together. The GEM foil glued to the bottom of the gas-tight enclosure is well visible.

A custom and very compact DAQ system to generate and handle command signals to/from the chip (implemented on Altera FPGA Cyclone EP1C240), to read and digitally convert the analog data (ADS5270TI Flash ADC) and to storage them, temporarily, on a static RAM, has been developed. By using the RISC processor NIOS II, embedded on Altera FPGA, and the self-triggering functionality of the chip, it is possible, immediately after the event is read-out, to acquire the pedestals of the pixels in the same *chip-defined* event window (region of interest). The pedestals can be read one or more times (user-defined). The average of the pedestal readings is used to transfer pedestal subtracted data to the off-line analysis system. This mode of operation has the great advantage of allowing the real time control of the data quality and to cancel any effect of time drift of the pedestal values or other temperature or environmental effects. The disadvantages are a slight increase of the channel noise (at maximum a factor √2, for the case of 1 pedestal reading only) and an increase of the event read-out time. For most of the applications we envisage, this will not be a real problem given the very large signal to noise ratio (well above 100) and the very fast operation in window mode.

Nonetheless, the acquisition of a set of pedestal values for all the 105k channels at the beginning or at the end of a data taking run in standard mode is still possible. The instrument control and data acquisition is done through a VI graphic interface developed in LAbVIEW. The VI performs a bidirectional communication through a 100Mbps TCP connection between the DAQ and a portable PC running Windows XP.

Fig.5 shows a snapshot of the Control Panel of the VI.

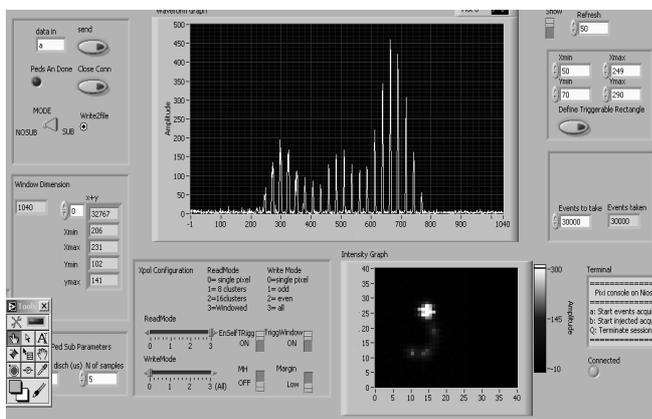

Fig. 5 – Snapshot of the Control Panel of the acquisition system. Together with the sets of user-selectable commands (chip configuration, self-trigger enable, on board pedestal subtraction,..) the canvas shows the display of one event as one-dimensional pixel charge distribution (ADC counts vs. pixel number in the event window) and the relative 2D image displayed in real time.

## 3. Laboratory tests and single photon operation

The detector has been tested in both configurations, *semitransparent* and *reflective*; in particular the capability to operate in single photon detection has been studied.

Data were acquired in *self-trigger* configuration and the detector was illuminated with a UV lamp operating in DC mode. The gas filling utilized is a mixture of 50% Neon - 50% DME.

The use of a GEM with a much finer pitch than usual (pitch and thickness have now equal size) that matches the 50μm read-out pitch, has allowed not only to push forward the 2D reconstruction capability of the device but also to obtain high gas gain with reduced voltages across the GEM. Fig.6 shows the measured effective gain as a function of the GEM voltage, with the typical exponential dependence, as derived from t the charge distributions of fig. 7.

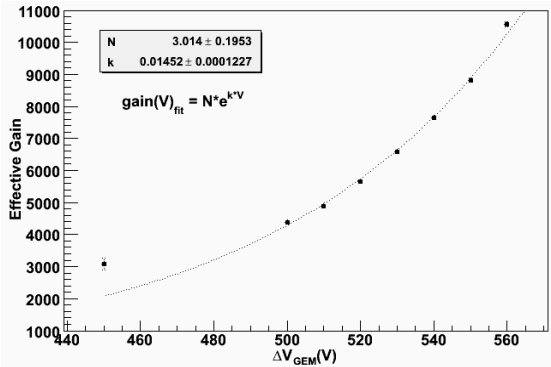

Fig. 6 - Effective gas gain vs. the voltage across the GEM obtained in the semitransparent photocathode configuration. The first point is out of the fit because of threshold effects at low gain.

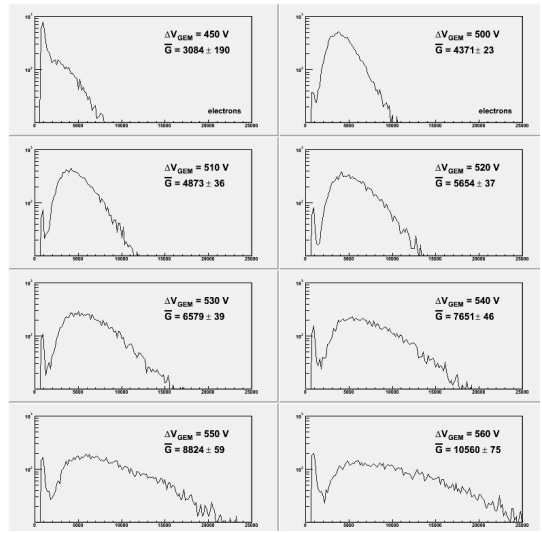

Fig. 7 - Charge distributions for different voltages across the GEM. The average effective gain obtained with a Polya fit to the distributions is reported.

The gain values reported in fig. 6 have been obtained by fitting (fig. 8) the corresponding distribution with a Polya function of the type:

$$P\left(\frac{G}{G_0}\right) = \frac{N m}{\Gamma(m)}\left(m \frac{G}{G_0}\right)^{m-1} e^{-m \frac{G}{G_0}}$$

where $G_0$ is the average effective gas gain.

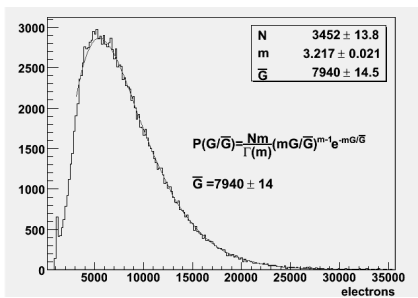

Fig. 8 - Typical charge distribution obtained with $\Delta V_{GEM}$ ~540 V in Ne50%-DME50%.

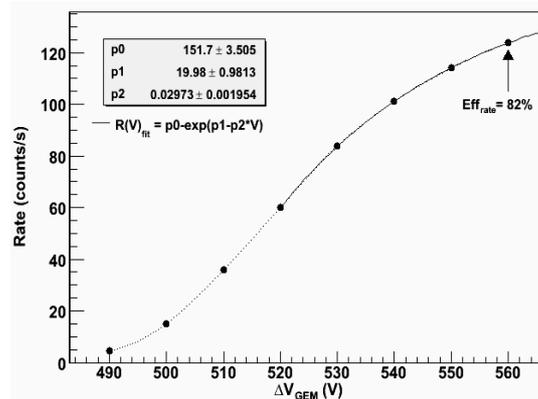

Fig. 9 - The dependence of the count rate with respect to the voltage applied to the GEM allows to estimate the detection efficiency of the device.

The detection efficiency of the device has been evaluated by measuring the trigger rate as a function of $\Delta V_{GEM}$ (fig. 9). From an exponential fit to the curve an efficiency of ~82% at 560V across the GEM has been obtained.

The characteristics of the detector of very low noise high gain and read-out granularity, allows to operate in single photon detection and, under this condition, a very high intrinsic spatial resolution has been measured.

With the UV light intensity kept sufficiently low, the device can detect one photon at a time, each producing a single electron.

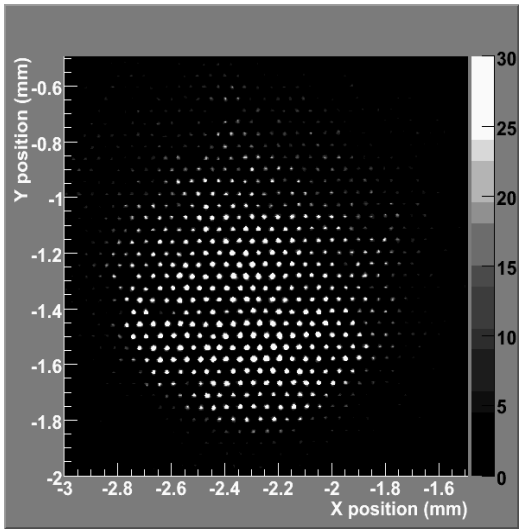

This electron drifts into a single GEM hole where it knocks out further electrons from atoms in an avalanche effect The avalanche started from the single electron is then extracted and collected onto the fine-pitch pixel CMOS analogue chip which provides a direct reading of the GEM charge multipliers. If the resolution is good and the noise is low, the centre of gravity corresponds to the centre of the GEM hole.

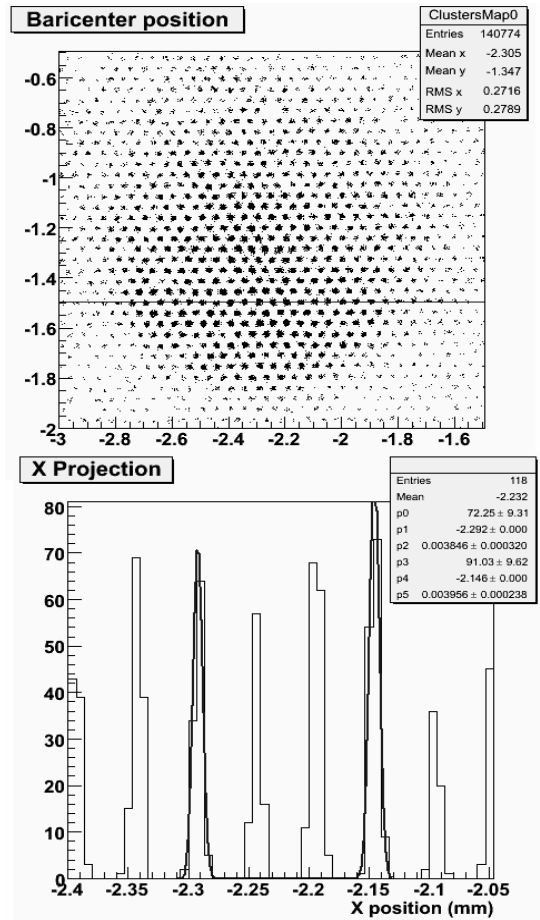

Fig. 10 - Cumulative map of hundreds thousands events from the UV light conversion in the semitransparent CsI photocathode of the detector producing a kind of *self-portrait* of the GEM amplification structure. The white spots correspond to the

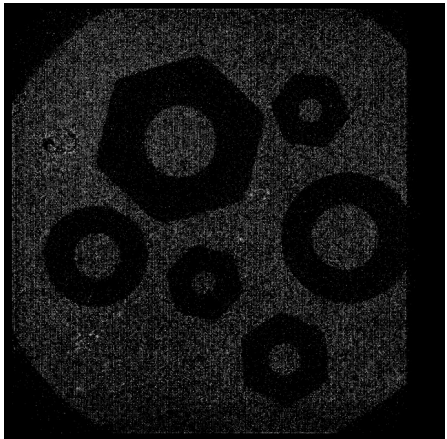

Fig. 12 - Reconstructed image of a sets of small mechanical components obtained by illuminating the detector from above with the UV lamp. The smallest object is a nut with internal diameter of only 1 mm.

Fig. 11 - An horizontal cut through the bi-dimensional map of the baricenter positions allows to estimate the intrinsic resolution of the detector. The peak to peak distance corresponds to the GEM pitch. A gaussian fit of the gives a width (rms) of less than 4 μm.

A cumulative map of several hundreds thousands of such events produces a kind of *self-portrait* of the GEM amplification structure with individual dots only 50 μm apart (fig. 10). A cut through the GEM holes of the 2D image, allows to estimate the intrinsic resolution of the read-out system, in response to a single primary electron, of only 4 μm (fig. 11). The picture shown in fig. 10 has been obtained with the CsI photocathode in semitransparent configuration, but similar, highly resolved, images have been obtained in reflective mode, as shown in fig. 12 where a set of small washers and nuts, the smallest one being an hexagonal nut with 1 mm internal diameter, has been imaged by illuminating the detector from above with the UV lamp.

At least three facts demonstrate that the detector can operate in single photon detection:
- after a first strong attenuation of the light flux by filtering, no significant change of the mean value and of the shape of the charge distribution has been observed by reducing further the trigger rate; at this point a final reduction of the flux of more than a factor five has been performed by adding UV filters and by moving the UV lamp far away from the entrance window.
- even at higher rate, double events, recognizable in the one dimensional plot of the pixel charge (fig. 13), are very rare and clearly distinguishable; in the 2D reconstruction they are in fact treated as two single photons. The 2D cumulative histogram of fig. 14 shows the first 50 triggers registered by the detector. All the 50 reconstructed photoelectron clusters are well separated.

- the spots corresponding to the GEM holes, in the baricenters distribution of fig. 11 are very well resolved; only the detection of a single photon cluster can lead to such a resolving power. Detecting two or more unresolved photons would make impossible to associate the baricenter with a GEM hole.

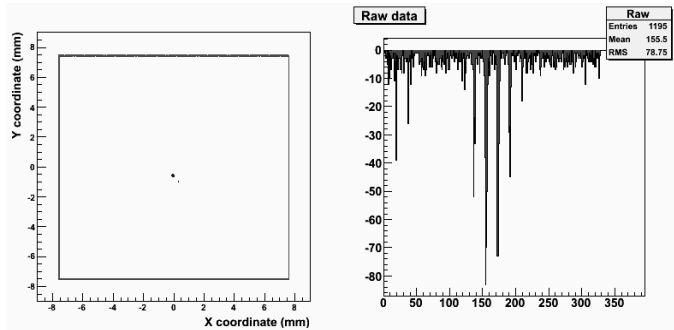

Fig. 13 – Example of double event. The two clusters, reconstructed in the 2D image, are also visible in the histogram on the right panel, which represents the collected charge of the pixels in the trigger window ordered according to the serialized read-out scheme. The smallest cluster in the 2D image appears in the left side of the histogram. The reduced amplitude of this cluster is due to the fact that this second photont comes out of time with respect to the main event which originates the trigger. It is worth to note that even if two events are registered within the same trigger window they are well distinguished in the bi-dimensional reconstruction.

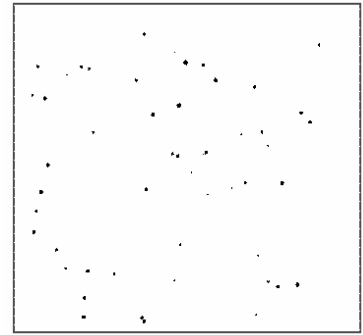

Fig. 14 - 2D reconstructed image of the first 50 triggers registered by illuminating the detector with the UV

## 4. Conclusions

A UV photo-detector based on a semitransparent or reflective CsI photocathode, followed by a Gas Electron Multiplier foil and a large area, custom, analog CMOS VLSI ASIC has been reported. The detector has shown excellent imaging capability and single electron sensitivity. The high granularity and low noise of the read-out electrode allows to reconstruct with a 4 $\mu$m resolution the centroid of the single electron avalanche. The position resolution of the device is, at the moment, limited by the 50 $\mu$m pitch of the GEM foil.

The main problem of such device is the control on the ion back-flow which can limit the lifetime of the detector, but various possible solutions exist and are now under development.

The performance obtained with this device allow to state that the class of Gas Pixel Detectors is now a mature technology with a level of integration, compactness, operational simplicity and resolving power up to now typical of solid state detectors only. It is also important to remark that this kind of device is very robust and after several months of intensive operation, no die or single pixel has been lost for electrostatic or GEM discharges, or for any other reason.